\documentclass[showpacs,aps,graphicx,onecolumn,10pt]{revtex4-2}
\usepackage{amsmath}
\usepackage{mathrsfs}
\usepackage{amsfonts}
\usepackage{amssymb}
\usepackage{graphicx}
\usepackage{caption}
\usepackage{subfigure}
\usepackage{eufrak}
\usepackage{multirow}
\usepackage{float}
\usepackage[colorlinks,linkcolor=blue,anchorcolor=blue,citecolor=blue]{hyperref}
\usepackage{soul,color,xcolor}

\begin{document}

\title{Quantum circuit optimization for arbitrary high-dimensional bipartite quantum computation}

\author{Gui-Long Jiang\textsuperscript{1,2} and Hai-Rui Wei\textsuperscript{1,}}
\email[]{hrwei@ustb.edu.cn}
\address{\textsuperscript{\rm1} School of Mathematics and Physics, University of Science and Technology Beijing, Beijing 100083, China\\
\textsuperscript{\rm2} School of Mathematics, Harbin Institute of Technology, Harbin 150001, China}

\begin{abstract}
Implementation of high-dimensional (HD) quantum gates shows very promising perspectives for HD quantum computation.
A bipartite quantum system with arbitrary dimensions $n$ and $m$ is termed a quNit-quMit.
Here we propose a synthesis scheme to construct the quantum circuit for general quNit-quMit gates with controlled increment (CINC) gates and local gates.
This shows that CINC gates combined with local gates form a universal gate set for HD quantum computation.
An upper bound of $O(n^2)$ CINC gates is achieved for arbitrary quNit-quMit gate implementation in the proposed scheme, which is the best known result.
Especially for the controlled quNit-quMit gates, our scheme requires only 2 CINC gates, whereas the previous scheme required $2n$.
\end{abstract}


\maketitle

\section{Introduction}\label{sec1}

A high-dimensional (HD) quantum system, called qudit ($d$-level with $d>2$), gradually exhibits remarkable advantages over binary systems in quantum information processing, due to its higher information capacity \cite{capacity}, better security against eavesdropping \cite{cryptography,noise-resistance}, and boosted algorithmic efficiency \cite{high-algorithm1,high-algorithm2,high-algorithm3}.
The physical qudit platforms have been naturally achieved in photonic systems \cite{Photonic,HD-GHZ-Photon}, continuous spin systems \cite{spin}, ion traps \cite{HD-entanglement,ion}, and superconducting circuits \cite{superconduct2,superconduct3,superconduct4}.
Up to now, qudit-based quantum communication tasks have been studied both theoretically and experimentally, including quantum key distribution \cite{QKD1,QKD3,QKD4}, quantum teleportation \cite{high-teleportation1,high-teleportation2,high-teleportation3}, and quantum cryptography \cite{cryptography,qutrit-cryptography}.
In the field of qudit-based quantum computing, quantum algorithms \cite{high-algorithm1,high-algorithm3}, quantum error correction \cite{correction1,correction2}, and HD quantum gates \cite{SUM-gate,Gao2020,hybrid-gate,GXOR-gate,Pi8-gate,single-qudit,X-gate1,X-gate2,CSUM-gate,Liu2024} have been widely discussed.
By employing accessible qudits, qubit-based circuit size, depth (the number of time steps required for quantum operations), and complexity (the number of elementary gates required for the quantum computation) might be further reduced, and the experimental setup for realizing qubit gates can be greatly simplified \cite{Simplified,Simplified1,Li-Wen-Dong,wei2020,Compression}.

It is impractical to construct a different physical setup for the realization of each multi-qudit operation.
A natural idea is to decompose arbitrary quantum operations into a sequence of simple-to-perform quantum gates. 
A set of quantum gates is called a universal gate set if any quantum operation can be synthesized from the gates in the set.
It is well known that single-qubit gates and controlled-NOT (CNOT) gates together form a universal gate set for multi-qubit computing \cite{Barenco1995}.
For multi-qudit computing, the collection of single-qudit gates together with a two-qudit \emph{imprimitive} gate that maps a certain product state to an entangled state is a universal gate set \cite{Brylinski2001}.
Some imprimitive gates for qudits, such as controlled-double-NOT (CDNOT) gates \cite{Li-Wen-Dong}, generalized controlled $X$ (GCX) gates \cite{Di2013}, and controlled increment (CINC) gates \cite{Brennen2006}, have been used to construct quantum circuits for implementing any multi-qudit operation.
A bipartite imprimitive gate is more difficult to realize than a local gate, as it is generally more susceptible to environmental noise.
This motivates using imprimitive gate counts to quantify the cost of a quantum circuit.

Optimizing the circuit cost is important as it reduces the circuit operation time and the probability of gate errors occurring.
For $n$-qutrit (3-level) systems, the current least-cost scheme requires $\frac{41}{96}\cdot3^{2n}-4\cdot3^{n-1}-(\frac{n^2}{2}+\frac{n}{4}-\frac{29}{32})$ GCX and CINC gates to synthesize a general $n$-qutrit gate \cite{Multi-qutrit}.
However, this scheme uses two types of imprimitive gates.
For $n$-ququart (4-level) systems, Li \emph{et al.} \cite{Li-Wen-Dong} proposed a scheme to construct a quantum circuit of general $n$-ququart gates, where $5(4^{2(n-1)}-4^{n-1})$ CDNOT gates are required.
The theoretical lower bound of $[d^{2n}-n(d^{2}-1)-1]/[4(d-1)]$ GCX gates for implementing a general $n$-qudit gate was derived by Di \emph{et al.} \cite{Di2015}.
However, no existing synthesis scheme has achieved the lower bound.
Several techniques of matrix decomposition are used to construct quantum circuits, such as quantum Shannon decomposition (QSD) \cite{Li-Wen-Dong,Di2013}, QR decomposition \cite{Brennen2006}, cosine-sine decomposition (CSD) \cite{Nakajima2009,Di2015}, spectral decomposition \cite{Bullock2005}, and Cartan decomposition \cite{Multi-qutrit,Cartan2023}.
In these synthesis schemes \cite{Li-Wen-Dong,Brennen2006,Multi-qutrit,Nakajima2009,Di2013}, any two-qudit gate is first decomposed into a sequence of the given imprimitive gates and single-qudit gates, and then any $n$-qudit gate is decomposed into two-qudit and single-qudit gates.
Thus, a quantum circuit for implementing general two-qudit gates is important because its cost significantly affects the overall circuit cost.
For qubits, the minimal cost of quantum circuits for implementing general two-qubit gates is 3 CNOTs \cite{2-qubit}.
For qudits, the minimal circuit cost for implementing general two-qudit gates remains an open challenge.

Here we focus on designing the quantum circuit for general quNit-quMit unitary operations with arbitrary dimensions $n$ and $m$.
In the paper, we choose a set that includes all single-qudit gates and a CINC gate as a universal gate set.
We first accurately implement a quNit-quMit controlled unitary gate using local gates and two CINC gates, and a quantum circuit for HD uniformly controlled unitary gates is constructed in section \ref{sec3}.
Subsequently, using CSD, we propose a synthesis scheme to construct the quantum circuit for general quNit-quMit gates in terms of CINC gates and local gates in section \ref{sec4}.
In contrast to the synthesis schemes using GCX \cite{Di2013,Di2015,Multi-qutrit}, our approach requires only a single type of CINC gate.
The controlled systems of all CINC gates in the quantum circuit are located on the same subsystem.
The number of universal gates required for our HD quantum circuits is the lowest currently known (see table \ref{Table1}), with detailed calculations provided in section \ref{sec5}.
The proposed synthesis scheme is not dependent on the physical platform.
As long as a physical system can realize CINC gates and local gates, the arbitrary HD quantum computation can be achieved according to our scheme.

\section{The concepts of high-dimensional quantum gates}\label{sec2}

\subsection{HD quantum gates on single systems}\label{sec2.1}

Let $\mathcal{H}_{n}$ be an $n$-dimensional quantum state space and $\{|1\rangle,\ldots,|n\rangle\}$ be an orthonormal basis of $\mathcal{H}_{n}$. Throughout the paper, the letters $i$, $j$, $k$, and $l$ denote positive integers.
The identity operator $\sum_{i=1}^{n}|i\rangle\langle i|$ on $\mathcal{H}_{n}$ is denoted by $I_{n}$.
For $1\leq i<j\leq n$, define three types of $n\times n$ Hermitian matrices as follows:
\begin{eqnarray}\label{eq1}
\sigma_{z_n}^{ij}=|i\rangle\langle i|-|j\rangle\langle j|,\;\;
\sigma_{x_n}^{ij}=|i\rangle\langle j|+|j\rangle\langle i|,\;\;
\sigma_{y_n}^{ij}=-\textrm{i}|i\rangle\langle j|+\textrm{i}|j\rangle\langle i|.
\end{eqnarray}

A $(i,j;\varphi_{n},\theta)$-rotation gate on $\mathcal{H}_{n}$ is defined as \cite{Di2013}
\begin{eqnarray}\label{eq2}
R_{\varphi_{n}}^{ij}(\theta)=\textrm{exp}(-\textrm{i}\frac{\theta}{2}\sigma_{\varphi_n}^{ij}),
\end{eqnarray}
where $1\leq i<j\leq n$, $\varphi\in \{x,y,z\}$, and the rotational parameter $\theta\in[0,2\pi]$.
It is a natural generalization of $\varphi$-axis rotation gates on a single qubit \cite{Shende2006}, as the action of $R_{\varphi_{n}}^{ij}(\theta)$ is to perform a $\varphi$-axis rotation gate with a rotational parameter $\theta$ on the two-dimensional state space spanned by $\{|i\rangle$, $|j\rangle\}$.
In addition, the $(i,j;\varphi_{n},\theta)$-rotation gate can be physically implemented by using Mach-Zehnder interferometers \cite{MZI1994,MZI2016}.

A generalized Pauli-$X$  operator on $\mathcal{H}_{n}$ is defined as \cite{single-qudit}
\begin{eqnarray}\label{eq3}
X_{n}=|1\rangle\langle n| +\sum^{n-1}_{i=1}|i+1\rangle\langle i|.
\end{eqnarray}
When $n=2$, the operator above is equal to the Pauli-$X$ (NOT) gate on a single qubit. 
The operator $X_{n}$ is also known as an increment gate, and $n-1$ powers of $X_{n}$ is equal to its conjugate transpose, i.e., $X^{\dag}_{n}=X^{n-1}_{n}$.
Hereafter, $\dagger$ denotes the conjugate transpose operation.
We define a unitary operator $T_{n}$ on $\mathcal{H}_{n}$ as
\begin{eqnarray}\label{eq4}
T_{n}=|1\rangle\langle 1| +\sum^{n}_{i=2}|i\rangle\langle n+2-i|.
\end{eqnarray}
It holds that 
\begin{eqnarray}\label{eq5}
T_{n}^{\dag}=T_{n},\;\;
T_{n} \cdot X_{n} \cdot T_{n}=X^{\dag}_{n}.
\end{eqnarray}

\subsection{HD quantum gates on bipartite systems}\label{sec2.2}
We write $\mathcal{H}_{n}^{1}\otimes\mathcal{H}_{m}^{2}$ to denote a bipartite quantum system composed of an $n$-dimensional state space $\mathcal{H}_{n}^{1}$ and a $m$-dimensional space $\mathcal{H}_{m}^{2}$.
Let $\{|i\rangle_{1}|1\leq i \leq n\}$ and $\{|j\rangle_{2}|1\leq j \leq m\}$ be orthonormal bases of $\mathcal{H}_{n}^{1}$ and $\mathcal{H}_{m}^{2}$, respectively.
The subscripts of $|i\rangle_{1}$ and $|i\rangle_{2}$ are omitted, and the notation $|i\rangle$ is used when the context clearly distinguishes between the first and second quantum systems.

\begin{figure} [htbp]
  \centering
  \includegraphics[width=14.5cm]{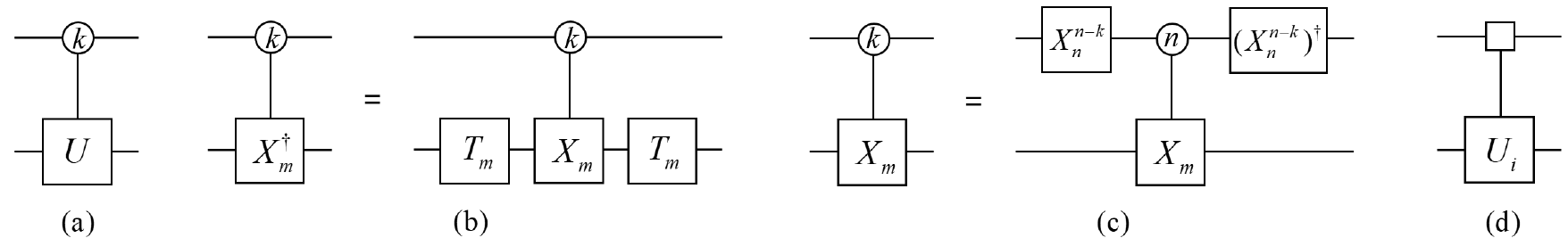}
  \caption{Quantum circuit symbols. (a) Controlled unitary gate with the control system state $|k\rangle_{1}$. The circle represents the control system. (b) Relation between $\textrm{C}_{k}(X_{m}^{\dag})$ and $\textrm{C}_{k}(X_{m})$. (c) Relation between $\textrm{C}_{n}(X_{m})$ and $\textrm{C}_{k}(X_{m})$. (d) Uniformly controlled unitary gate with the control system $\mathcal{H}_{n}^{1}$. The square ($\Box$) denotes the control system.}
  \label{Fig.1}
\end{figure}

An operator on $\mathcal{H}_{n}^{1}\otimes\mathcal{H}_{m}^{2}$ is said to be a controlled unitary gate with the control system state $|k\rangle_{1}$, if its action is to perform a specific unitary operator on the target system $\mathcal{H}_{m}^{2}$ when the state of $\mathcal{H}_{n}^{1}$ is $|k\rangle_{1}$; otherwise, it does nothing.
The mathematical representation of the controlled unitary gate with the control system state $|k\rangle_{1}$ can be written as
\begin{eqnarray}\label{eq6}
\textrm{C}_{k}(U)=|k\rangle\langle k| \otimes U+\sum^{n}_{i =1,i \neq k}|i\rangle\langle i| \otimes I_m,
\end{eqnarray}
where $U$ is a unitary operator on $\mathcal{H}_{m}^{2}$ and $I_m$ is the identity operator on $\mathcal{H}_{m}^{2}$.
Its quantum circuit is given in figure \ref{Fig.1}(a).
Equation \eqref{eq6} implies the following properties
\begin{eqnarray}\label{eq7}
\begin{split}
&\textrm{C}_{k}(U^{\dag})=(\textrm{C}_{k}(U))^{\dag},\\[1mm]
&\textrm{C}_{k}(UV)=\textrm{C}_{k}(U) \cdot \textrm{C}_{k}(V),\\[1mm]
&\textrm{C}_{k}(U)\cdot \textrm{C}_{l}(V)=\textrm{C}_{l}(V)\cdot \textrm{C}_{k}(U),\\[1mm]
&\textrm{C}_{k}(VUV^{\dag}) = I_{n}\otimes V \cdot \textrm{C}_{k}(U) \cdot I_{n}\otimes V^{\dag}.
\end{split}
\end{eqnarray}
for every $k\neq l\in\{1,\ldots,n\}$ and any two unitary operators $U$ and $V$ on $\mathcal{H}_{m}^{2}$.

Particularly, when $U=X_{m}$ in equation \eqref{eq6}, it is defined as a controlled-$X_{m}$ (i.e., CINC) gate \cite{Liu2024}
\begin{eqnarray}\label{eq8}
\textrm{C}_{k}(X_{m})=|k\rangle\langle k| \otimes X_{m}+\sum^{n}_{i =1,i \neq k}|i\rangle\langle i| \otimes I_m.
\end{eqnarray}
From equation \eqref{eq5} and equation \eqref{eq7}, it holds that
\begin{eqnarray}\label{eq9}
\begin{split}
\textrm{C}_{k}(X_{m}^{\dag}) &= \textrm{C}_{k}(T_{m} \cdot X_{m} \cdot T_{m})\\[1mm]
&=I_{n} \otimes T_{m}\cdot \textrm{C}_{k}(X_{m}) \cdot I_{n} \otimes T_{m}.
\end{split}
\end{eqnarray}
Equation \eqref{eq9} means that $\textrm{C}_{k}(X_{m}^{\dag})$ is equivalent to $\textrm{C}_{k}(X_{m})$ up to two local unitary operators $T_{m}$, as shown in figure \ref{Fig.1}(b).
Moreover, $\textrm{C}_{k}(X_{m})$ can be transformed into $\textrm{C}_{n}(X_{m})$ by local gates on $\mathcal{H}_{n}^{1}$, as shown in figure \ref{Fig.1}(c).

When the control system is a qubit (i.e., $\mathcal{H}_{2}^{1}\otimes\mathcal{H}_{m}^{2}$), $\textrm{C}_{1}(X_{m})$ or $\textrm{C}_{2}(X_{m})$ can be experimentally realized effectively using only linear optical elements and the polarization degree of freedom of photons \cite{Liu2024,Meng2022}.
However, for $\mathcal{H}_{n}^{1}\otimes\mathcal{H}_{m}^{2}$ where $n>2$, it is extremely difficult to implement $\textrm{C}_{k}(X_{m})$ solely using the photonic polarization, as photonic polarization is inherently insufficient to encode more than a two-level system.
Although this is not directly relevant to the main results of this paper, for the sake of completeness, we provide an experimental implementation of $\textrm{C}_{n}(X_{m})$ where $n>2$ in Appendix \ref{Appendix.A}.

The operator $\textrm{C}_{k}(U)$ may be generalized as follows.
Let $\{U_{1},\ldots,U_{n}\}$ be a unitary operator set on $\mathcal{H}_{m}^{2}$.
A uniformly controlled unitary gate with the control system $\mathcal{H}_{n}^{1}$ is defined as 
\begin{eqnarray}\label{eq10}
\begin{split}
\sum_{i=1}^{n}|i\rangle\langle i| \otimes U_{i}=\left[
  \begin{array}{cccc}
    U_{1} & \mathbf{0} & \cdots & \mathbf{0}  \\
    \mathbf{0} & U_{2} & \cdots & \mathbf{0}  \\
    \vdots & \vdots & \ddots & \vdots  \\
    \mathbf{0} & \mathbf{0} & \cdots & U_{n}  \\
  \end{array}
\right].
\end{split}
\end{eqnarray}
The action of such a gate on $\mathcal{H}_{n}^{1}\otimes\mathcal{H}_{m}^{2}$ is to perform the unitary operator $U_{i}$ on the target system $\mathcal{H}_{m}^{2}$ when the state of $\mathcal{H}_{n}^{1}$ is $|i\rangle_{1}$.
The quantum circuit representation of such a gate is shown in figure \ref{Fig.1}(d).
Equivalently, the uniformly controlled unitary gate with control system $\mathcal{H}_{m}^{2}$ is defined in the same manner as described above.
For example, given a real diagonal matrix $D_{m}=\sum_{k=1}^{m}\theta_{k}|k\rangle\langle k|$, we find that
\begin{eqnarray}\label{eq11}
\begin{split}
\textrm{exp}(-\textrm{i}\sigma_{\varphi_n}^{ij} \otimes D_m)
&=\textrm{exp}(-\sum_{k=1}^{m}\theta_{k}\textrm{i}\sigma_{\varphi_n}^{ij} \otimes |k\rangle\langle k|)\\
&=\sum_{k=1}^{m}\textrm{exp}(-\theta_{k}\textrm{i}\sigma_{\varphi_n}^{ij} \otimes |k\rangle\langle k|)\\
&=\sum_{k=1}^{m}\textrm{exp}(-\theta_{k}\textrm{i}\sigma_{\varphi_n}^{ij}) \otimes |k\rangle\langle k|\\
&=\sum_{k=1}^{m}R^{ij}_{\varphi_{n}}(2\theta_{k}) \otimes |k\rangle\langle k|.
\end{split}
\end{eqnarray}
The second equality follows from the commutativity of $\sigma_{\varphi_n}^{ij} \otimes |k\rangle\langle k|$ and $\sigma_{\varphi_n}^{ij} \otimes |l\rangle\langle l|$ when $k\neq l$, and the last equality is due to equation \eqref{eq2}.
Hence, $\textrm{exp}(-\textrm{i}\sigma_{\varphi_n}^{ij} \otimes D_m)$ is a uniformly controlled $R^{ij}_{\varphi_{n}}$ gate with the control system $\mathcal{H}_{m}^{2}$. 
In the subsequent discussion, we denote $R^{ij}_{\varphi_{n}}$ simply as $R_{\varphi_{n}}$ if the indices $ij$ are unimportant in the context.

\section{Quantum circuits for quNit-quMit controlled unitary operations}\label{sec3}
\subsection{Quantum circuits for controlled unitary gates}\label{sec3.1}
\begin{figure} [htbp]
  \centering
  \includegraphics[width=13cm]{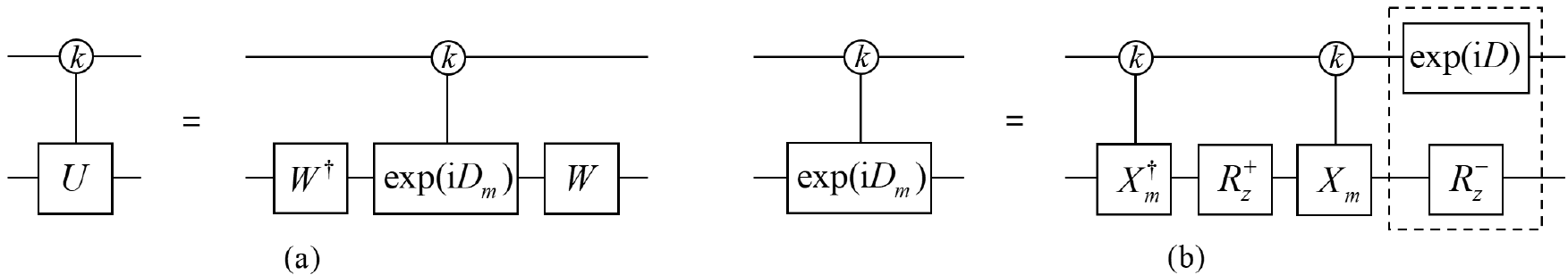}
  \caption{(a) Equivalent quantum circuit for $\textrm{C}_{k}(U)$ composed of a controlled diagonal gate and two local operations. (b) Equivalent quantum circuit for $\textrm{C}_{k}(e^{\textrm{i}D_{m}})$ based on equation \eqref{eq19}. Since $\textrm{C}_{k}(X_{m}) \cdot I_{n} \otimes R_{z}^{+} \cdot \textrm{C}_{k}(X_{m}^{\dag})$ in equation \eqref{eq19} is a diagonal matrix, the local operations in the dotted box can be moved to the left of $\textrm{C}_{k}(X^{\dag}_{m})$.}
  \label{Fig.2}
\end{figure}

Ref. \cite{Brennen2006} uses $2n$ $\textrm{C}_{k}(X_{m})$ gates to implement $\textrm{C}_{k}(U)$.
Now we use local operations and only two $\textrm{C}_{k}(X_{m})$ gates to implement $\textrm{C}_{k}(U)$.
Since the unitary matrix is diagonalizable, it may be assumed that $U=W \cdot \textrm{exp}(\textrm{i}D_{m}) \cdot W^{\dag}$, where $W$ is a unitary matrix and $D_{m}=\sum_{k=1}^{m}\theta_{k}|k\rangle\langle k|$ is a real diagonal matrix.
From equation \eqref{eq7}, $\textrm{C}_{k}(U)$ is equivalent to a controlled diagonal gate 
\begin{eqnarray}\label{eq12}
\begin{split}
\textrm{C}_{k}(e^{\textrm{i}D_{m}})&=\textrm{exp}(|k\rangle\langle k| \otimes \textrm{i}D_m) \\[1mm]
&=|k\rangle\langle k| \otimes \textrm{exp}(\textrm{i}D_{m})+\sum^{n}_{i =1,i \neq k}|i\rangle\langle i| \otimes I_m,
\end{split}
\end{eqnarray}
up to two local operations, as shown in figure \ref{Fig.2}(a).
It therefore suffices to focus on the decomposition of $\textrm{C}_{k}(e^{\textrm{i}D_{m}})$.

We first express $D_{m}=\textrm{diag}\{\theta_{1},\ldots,\theta_{m}\}$ as a linear combination of a set of special diagonal matrices.
Let diagonal matrices $E_{1}=I_{m}$, and for $i \in \{2,\ldots,m\}$,
\begin{eqnarray}\label{eq13}
\begin{split}
E_{i}&=\sigma_{z_m}^{i-1,i}-X_{m} \cdot \sigma_{z_m}^{i-1,i} \cdot X^{\dagger}_{m}\\[1mm]
&=
\left\{
\begin{array}{cll}
\sigma_{z_m}^{i-1,i} - \sigma_{z_m}^{i,i+1},   & \; \mathrm{if}\;  i \in \{2,\ldots,m-1\},  &     \\[2mm]
\sigma_{z_m}^{m-1,m} + \sigma_{z_m}^{1,m},     & \; \mathrm{if}\;  i = m.  &   \\
\end{array}
\right.
\end{split}
\end{eqnarray}
Since the set $\{I_{m}, \sigma_{z_m}^{1,2},\ldots,\sigma_{z_m}^{m-1,m}\}$ is linearly independent and $\sigma_{z_m}^{1,m}=\sum^{m}_{i=2}\sigma_{z_m}^{i-1,i}$, one can easily deduce that $\{E_{1},\ldots,E_{m}\}$ is also linearly independent by the fact that
\begin{eqnarray}\label{eq14}
\begin{split}
\sigma_{z_m}^{i-1,i}=\sum^{m}_{k=i}E_{k}-\sigma_{z_m}^{1,m},\;\;
\sigma_{z_m}^{1,m}=\frac{\sum^{m}_{l=2}(l-1)E_{l}}{m}.
\end{split}
\end{eqnarray}
Hence, there exist unique real numbers $x_{1},\ldots,x_{m}$ such that $D_{m}=\sum_{i=1}^{m}x_{i}E_{i}$.
Furthermore, one has that
\begin{eqnarray}\label{eq15}
\begin{split}
\exp(\textrm{i}D_{m})=&\prod^{m}_{i=1}\exp(\textrm{i}x_{i}E_{i})\\
=&e^{\textrm{i}x_1}\prod^{m}_{i=2}\exp\big[\textrm{i}x_{i}(\sigma_{z_m}^{i-1,i}-X_{m} \cdot \sigma_{z_m}^{i-1,i} \cdot X^{\dagger}_{m})\big]\\
=&e^{\textrm{i}x_1}\prod^{m}_{i=2}[\exp(\textrm{i}x_{i}\sigma_{z_m}^{i-1,i}) \cdot \exp(-\textrm{i}x_{i} X_{m} \cdot \sigma_{z_m}^{i-1,i} \cdot X^{\dagger}_{m})]\\
=&e^{\textrm{i}x_1}\prod^{m}_{i=2}\exp(\textrm{i}x_{i}\sigma_{z_m}^{i-1,i}) \cdot \prod^{m}_{i=2}\exp(-\textrm{i}x_{i} X_{m} \cdot \sigma_{z_m}^{i-1,i} \cdot X^{\dagger}_{m})\\
=&e^{\textrm{i}x_1}\prod^{m}_{i=2}R_{z_m}^{i-1,i}(-2x_{i}) \cdot X_{m} \cdot \prod^{m}_{i=2}R_{z_m}^{i-1,i}(2x_{i}) \cdot X^{\dagger}_{m}.
\end{split}
\end{eqnarray}
In the derivation of equation \eqref{eq15}, the second equality follows from equation \eqref{eq13};
the third and fourth equalities follow the fact that $X_{m} \cdot \sigma_{z_m}^{i-1,i} \cdot X^{\dagger}_{m}$ is a diagonal matrix;
in the last equality we use equation \eqref{eq2} and the relation $\textrm{exp}(B\cdot A \cdot B^{-1})=B\cdot\textrm{exp}(A)\cdot B^{-1}$, where $B^{-1}$ denotes the inverse matrix of $B$.
Let $R_{z}^{\pm}=\prod^{m}_{i=2}R_{z_m}^{i-1,i}(\pm2x_{i})$. From equations \eqref{eq15} and \eqref{eq7}, we have
\begin{eqnarray}\label{eq16}
\textrm{C}_{k}(e^{\textrm{i}D_{m}})=\textrm{C}_{k}(e^{\textrm{i}x_1}I_{m}) \cdot \textrm{C}_{k}(R_{z}^{-} \cdot X_{m} \cdot R_{z}^{+}) \cdot \textrm{C}_{k}(X_{m}^{\dag}).
\end{eqnarray}

By the definition of $R_{z_{m}}^{ij}(\theta)$, it follows that $(R_{z_{m}}^{ij}(\theta))^{\dag}=R_{z_{m}}^{ij}(-\theta)$. Then, one has that
\begin{eqnarray}\label{eq17}
(R_{z}^{+})^{\dag} = R_{z}^{-}.
\end{eqnarray}
In addition, the first term $\textrm{C}_{k}(e^{\textrm{i}x_1}I_{m})$ in the left-hand side of equation \eqref{eq16} is essentially a local diagonal operator on $\mathcal{H}_{n}^{1}$ through the following equation
\begin{eqnarray}\label{eq18}
\begin{split}
\textrm{C}_{k}(e^{\textrm{i}x_1}I_{m})&=|k\rangle\langle k| \otimes e^{\textrm{i}x_1}I_{m}+\sum^{n}_{i =1,i \neq k}|i\rangle\langle i| \otimes I_m\\
&=\textrm{exp}(\textrm{i}D) \otimes I_{m}.
\end{split}
\end{eqnarray}
Here the diagonal matrix $D=x_{1}|k \rangle\langle k|$. From equations \eqref{eq17}, \eqref{eq7}, and \eqref{eq18}, we obtain 
\begin{eqnarray}\label{eq19}
\textrm{C}_{k}(e^{\textrm{i}D_{m}})=\textrm{exp}(\textrm{i}D) \otimes R_{z}^{-} \cdot \textrm{C}_{k}(X_{m}) 
\cdot I_{n} \otimes R_{z}^{+} \cdot \textrm{C}_{k}(X_{m}^{\dag}).
\end{eqnarray}

Figure \ref{Fig.2}(b) shows an equivalent quantum circuit of $\textrm{C}_{k}(e^{\textrm{i}D_{m}})$ based on equation \eqref{eq19}.
Here $\textrm{C}_{k}(X_{m}^{\dag})$ can be implemented by $\textrm{C}_{k}(X_{m})$ and local operations from figure \ref{Fig.1}(c).
Hence, combining with figure \ref{Fig.2}(a), local operations and two $\textrm{C}_{k}(X_{m})$ are sufficient to implement $\textrm{C}_{k}(U)$.

\subsection{Quantum circuits for uniformly controlled unitary gates}\label{sec3.2}
\begin{figure} [htbp]
  \centering
  \includegraphics[width=8.5cm]{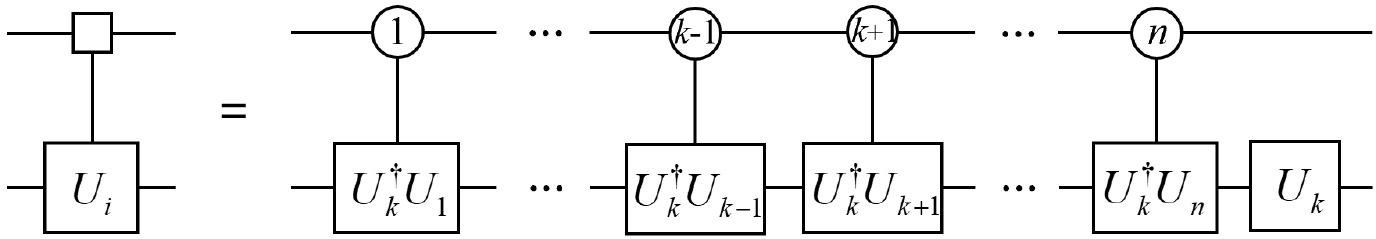}
  \caption{Equivalent quantum circuit for the uniformly controlled unitary gate $\sum_{i=1}^{n}|i\rangle\langle i| \otimes U_{i}$. Here $k\in\{1,\ldots,n\}$. From equation \eqref{eq7}, the order of the $n-1$ controlled unitary gates can be arranged arbitrarily.}
  \label{Fig.3}
\end{figure}

Applying the above results, we can obtain a decomposition of the uniformly controlled unitary gate $\sum_{i=1}^{n}|i\rangle\langle i| \otimes U_{i}$ with the control system $\mathcal{H}_{n}^{1}$.
For every $k\in\{1,\ldots,n\}$, we have
\begin{eqnarray}\label{eq20}
\begin{split}
\sum_{i=1}^{n}|i\rangle\langle i| \otimes U_{i}=&\prod^{n}_{i=1}\textrm{C}_{i}(U_{i})\\ 
=& I_{n}\otimes U_{k} \cdot I_{n}\otimes U_{k}^{\dag} \cdot \prod^{n}_{i=1}\textrm{C}_{i}(U_{i}) \\
=& I_{n}\otimes U_{k} \cdot \prod^{n}_{i=1}\textrm{C}_{i}(U_{k}^{\dag})  \cdot \prod^{n}_{i=1}\textrm{C}_{i}(U_{i})  \\
=& I_{n}\otimes U_{k} \cdot \prod^{n}_{i=1,i\neq k}\textrm{C}_{i}(U_{k}^{\dag}U_{i}),
\end{split}
\end{eqnarray}
where the last equality follows from the third and fourth formulas in equation \eqref{eq7}.
Hence, $\sum_{i=1}^{n}|i\rangle\langle i| \otimes U_{i}$ is equivalent to $(n-1)$ control unitary gates $\{\textrm{C}_{i}(U_{k}^{\dag}U_{i})|1 \leq i \leq n,\;i\neq k\}$ under the action of one local operation, as shown in figure \ref{Fig.3}.
Together with figure \ref{Fig.2}, it is clear that a uniformly controlled unitary gate with control system $\mathcal{H}_{n}^{1}$ can be implemented by local gates and $2(n-1)$ controlled-$X_{m}$ gates.

\begin{figure} [htbp]
  \centering
  \includegraphics[width=13cm]{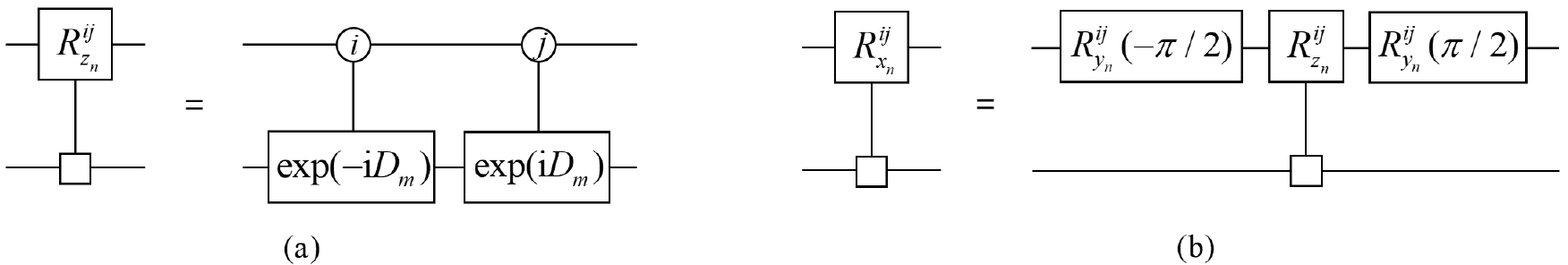} 
  \caption{(a) Equivalent quantum circuit for the uniformly controlled $R^{ij}_{z_{n}}$ gate with control system $\mathcal{H}_{m}^{2}$.
  (b) Equivalent quantum circuit for the uniformly controlled $R^{ij}_{x_{n}}$ gate with control system $\mathcal{H}_{m}^{2}$.}
  \label{Fig.4}
\end{figure}

Moreover, from equation \eqref{eq11}, it is known that $\textrm{exp}(-\textrm{i}\sigma_{z_n}^{ij}\otimes D_m)$ is a uniformly controlled $R^{ij}_{z_{n}}$ gate with control system $\mathcal{H}_{m}^{2}$, where $D_{m}=\sum_{k=1}^{m}\theta_{k}|k\rangle\langle k|$.
By the definition of $\sigma_{z_n}^{ij}$, we have 
\begin{eqnarray}\label{eq21}
\begin{split}
\textrm{exp}(-\textrm{i}\sigma_{z_n}^{ij} \otimes D_m)
=&\;\textrm{exp}[\textrm{i}(|j\rangle\langle j|-|i\rangle\langle i|) \otimes D_m]\\
=&\;\textrm{exp}(|j\rangle\langle j| \otimes \textrm{i}D_m) \cdot \textrm{exp}[|i\rangle\langle i| \otimes (-\textrm{i}D_m)]\\
=&\;\textrm{C}_{j}(e^{\textrm{i}D_{m}}) \cdot \textrm{C}_{i}(e^{-\textrm{i}D_{m}}).
\end{split}
\end{eqnarray}
Thus, such an operator can be decomposed into a product of two controlled diagonal gates $\textrm{C}_{i}(e^{-\textrm{i}D_{m}})$ and $\textrm{C}_{j}(e^{\textrm{i}D_{m}})$, as shown in figure \ref{Fig.4}(a).
Moreover, we find
\begin{eqnarray}\label{eq22}
\begin{split}
\sigma_{x_n}^{ij} =R_{y_{n}}^{ij}(\frac{\pi}{2}) \cdot\sigma_{z_n}^{ij}\cdot R_{y_{n}}^{ij}(-\frac{\pi}{2}).
\end{split}
\end{eqnarray}
Together with $(R_{y_{n}}^{ij}(\frac{\pi}{2}))^{\dagger}=R_{y_{n}}^{ij}(-\frac{\pi}{2})$, we have
\begin{eqnarray}\label{eq23}
\textrm{exp}(-\textrm{i}\sigma_{x_n}^{ij} \otimes D_m)
=R_{y_{n}}^{ij}(\frac{\pi}{2})\otimes I_m \cdot\textrm{exp}(-\textrm{i}\sigma_{z_n}^{ij} \otimes D_m)\cdot R_{y_{n}}^{ij}(-\frac{\pi}{2})\otimes I_m.
\end{eqnarray}
Based on equation \eqref{eq23}, one finds that a uniformly controlled $R^{ij}_{x_{n}}$ gate is equivalent to a uniformly controlled $R_{z_{n}}^{ij}$ gate up to local $R^{ij}_{y_{n}}$ gates, as shown in figure \ref{Fig.4}(b).
From figure \ref{Fig.2}(b), 4 controlled-$X_{m}$ gates are sufficient to implement a uniformly controlled $R^{ij}_{z_{n}}$  ($R^{ij}_{x_{n}}$) gate with control system $\mathcal{H}_{m}^{2}$.
In addition, when $n=3$, only 3 controlled-$X_{m}$ gates are sufficient, by applying figure 1(a) in Ref. \cite{Multi-qutrit}.

\section{Quantum circuits for general quNit-quMit gates}\label{sec4}
Next, we show that $\textrm{C}_{n}(X_{m})$ and local gates form a universal set for HD quantum computation.
The CSD provides an effective technique for synthesizing arbitrary multi-qubit quantum gates \cite{Shende2006} as well as general HD quantum gates \cite{Nakajima2009}.
In the following, we use CSD to give a recursive decomposition that breaks down general unitary operations on $\mathcal{H}_{n}^{1}\otimes\mathcal{H}_{m}^{2}$ into a product of uniformly controlled unitary gates with the control system $\mathcal{H}_{n}^{1}$ and uniformly controlled $R_{x_{n}}$ gates with the control system $\mathcal{H}_{m}^{2}$.

\subsection{Synthesis algorithm of general quNit-quMit gates}\label{sec4.1}
Let $U(n)$ denote the unitary group of $n\times n$ unitary matrices.
A general unitary operation on $\mathcal{H}_{n}^{1}\otimes\mathcal{H}_{m}^{2}$ can be recognized as an element of $U(nm)$.

\emph{Cosine–sine decomposition}.---Let $X \in U(nm)$ be a unitary matrix partitioned as
\begin{eqnarray}\label{eq24}
X=\left[
  \begin{array}{cc}
    X_{11} & X_{12}           \\
    X_{21} & X_{22}           \\
  \end{array}
\right],
\end{eqnarray}
where $X_{11}$ is a $\lfloor \frac{n}{2} \rfloor m \times \lfloor \frac{n}{2} \rfloor m$ matrix and $X_{22}$ is a $(n-\lfloor \frac{n}{2} \rfloor) m \times (n-\lfloor \frac{n}{2} \rfloor) m$ matrix.
For convenience, setting $n_{1}=\lfloor \frac{n}{2} \rfloor$ and $n_{2}=n-\lfloor \frac{n}{2} \rfloor$, we have $0<n_{1}\leq n_{2}<n$ and $n_{1}+n_{2}=n$.
The CSD \cite{Nakajima2009,Di2015,CSD} factorizes $X$ into a product of three matrices:
\begin{eqnarray}\label{eq25}
\begin{split}
X=&\left[
  \begin{array}{cc}
    U_{1} & \mathbf{0}           \\
    \mathbf{0} & \bar{U}_{2}           \\
  \end{array}
\right]
\cdot
\left[
  \begin{array}{ccc}
    C & -S &  \mathbf{0}   \\
    S & C  &  \mathbf{0}   \\
    \mathbf{0} & \mathbf{0} & I_{(n_{2}-n_{1})m}  \\
  \end{array}
\right]
\cdot
\left[
  \begin{array}{cc}
    U'_{1} & \mathbf{0}           \\
    \mathbf{0} & \bar{U}'_{2}           \\
  \end{array}
\right],
\end{split}
\end{eqnarray}
where $U_1,U'_1 \in U(n_1 m)$, $\bar{U}_2,\bar{U}'_2 \in U(n_2m)$, and $C,S$ are $n_1m \times n_1m$ real diagonal matrices satisfying $C^2+S^2=I_{n_1}$.

In Ref. \cite{Nakajima2009}, the second matrix in the right-hand side of equation \eqref{eq25} can be expressed as a product of uniformly controlled $R_{y_{n}}$ gates with the control system $\mathcal{H}_{m}^{2}$.
Next, we make a slight modification to equation \eqref{eq25} such that the second matrix becomes a product of uniformly controlled $R_{x_{n}}$ gates with the control system $\mathcal{H}_{m}^{2}$.

\emph{The first decomposition of} $U(nm)$.---From equation \eqref{eq25}, we can get a decomposition of $X\in U(nm)$ as
\begin{eqnarray}\label{eq26}
\begin{split}
X=U\cdot V \cdot U'
\end{split}
\end{eqnarray}
where
\begin{eqnarray}\label{eq27}
\begin{split}
V=\left[
  \begin{array}{ccc}
    C & -\textrm{i}S &  \mathbf{0}   \\
    -\textrm{i}S & C  &  \mathbf{0}   \\
    \mathbf{0} & \mathbf{0} & I_{(n_{2}-n_{1})m}  \\
  \end{array}
\right].
\end{split}
\end{eqnarray}
\begin{eqnarray}\label{eq28}
\begin{split}
U&=\left[
  \begin{array}{cc}
    U_{1} & \mathbf{0}           \\
    \mathbf{0} & U_{2}           \\
  \end{array}
\right],\;
U_{2}=\bar{U}_{2}
\cdot
\left[
  \begin{array}{ccc}
   \textrm{i}I_{n_{1}m}  &  \mathbf{0}   \\
   \mathbf{0} & I_{(n_{2}-n_{1})m}  \\
  \end{array}
\right],\\[2mm]
U'&=\left[
  \begin{array}{cc}
    U'_{1} & \mathbf{0}           \\
    \mathbf{0} & U'_{2}           \\
  \end{array}
\right],\;
U'_{2}=
\left[
  \begin{array}{cc}
     -\textrm{i}I_{n_{1}m}  &  \mathbf{0}   \\
     \mathbf{0} & I_{(n_{2}-n_{1})m}  \\
  \end{array}
\right]
\cdot
\bar{U}'_{2}.
\end{split}
\end{eqnarray}

We now explain that $V$ can be expressed as a product of uniformly controlled $R_{x_{n}}$ gates.
Without loss of generality, we may assume 
\begin{eqnarray}\label{eq29}
\begin{split}
C&=\textrm{diag}\{\cos\theta_{11},\ldots,\cos\theta_{1m},\cos\theta_{21},\ldots,\cos\theta_{n_1m}\},
\\[1mm]
S&=\textrm{diag}\{\sin\theta_{11},\ldots,\sin\theta_{1m},\sin\theta_{21},\ldots,\sin\theta_{n_1m}\}.
\end{split}
\end{eqnarray}
One can verify that
\begin{eqnarray}\label{eq30}
V=\textrm{exp}\big\{\sum_{i=1}^{n_1}-
\textrm{i}\sigma_{x_n}^{i,n_1+i}
\otimes D_m^{(i)}\big\}
=\prod_{i=1}^{n_1}
\textrm{exp}(-\textrm{i}\sigma_{x_n}^{i,n_1+i} \otimes D_m^{(i)}),
\end{eqnarray}
where the first equality is obtained by letting the diagonal matrix $D_m^{(i)}=\textrm{diag}\{\theta_{i1},\ldots,\theta_{im}\}$;
the second equality follows from the commutativity of two matrices $\sigma_{x_n}^{i,n_1+i}$ and $\sigma_{x_n}^{j,n_1+j}$.

\begin{figure}[htbp]
  \centering
  \includegraphics[width=6.6cm]{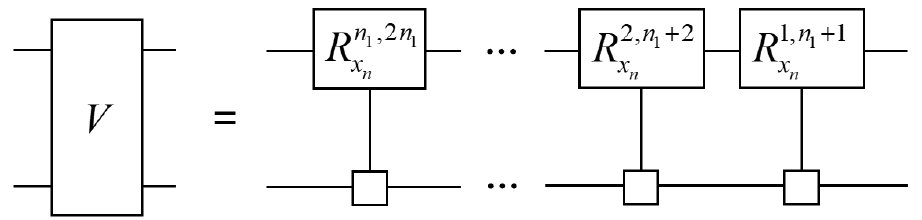}
  \caption{Quantum circuit for the non-local operation $V$ given by equation (\ref{eq30}). Here $n_1=\lfloor\frac{n}{2}\rfloor$.}\label{Fig.5}
\end{figure}

From equations \eqref{eq11} and \eqref{eq30}, $V$ is equivalent to $\lfloor\frac{n}{2}\rfloor$ uniformly controlled $R_{x_{n}}$ gates with the control system $\mathcal{H}_{m}^{2}$, as shown in figure \ref{Fig.5}.
Hence, combining with figure \ref{Fig.2}(b) and figure \ref{Fig.4}, local operations and $4\lfloor\frac{n}{2}\rfloor$ controlled-$X_{m}$ gates are sufficient to implement $V$ exactly.

From equations \eqref{eq25} and \eqref{eq28}, $U_1,U'_1 \in U(n_1 m)$ and $U_2,U'_2 \in U(n_2 m)$.
If $n_1=n_2=1$ (i.e., $n=2$), $U$ and $U'$ are uniformly controlled unitary gates with the control system $\mathcal{H}_{2}^{1}$, which can be synthesized by figure \ref{Fig.3} and figure \ref{Fig.2}.
In this case, the synthesis of $X$ is completed.
Otherwise, one can find that $U,U'\in U(n_1 m) \oplus U(n_2 m)$ can be further decomposed using equation \eqref{eq26} as follows.

\emph{The second decomposition of} $U(nm)$.---
Here we take $U$ as an example, and $U'$ will follow a similar discussion to $U$.
We first discuss the case where $n_1>1$.
Applying equation \eqref{eq26} to decompose $U_1$ and $U_2$, we get
\begin{eqnarray}\label{eq31}
\left[
  \begin{array}{cc}
    U_{1} & \mathbf{0}           \\
    \mathbf{0} & U_{2}           \\
  \end{array}
\right]=
\left[
  \begin{array}{cc}
    W_{1} & \mathbf{0}   \\
    \mathbf{0} & W_{2}   \\
  \end{array}
\right]
\cdot
\left[
  \begin{array}{cc}
    V_{1} & \mathbf{0}   \\
    \mathbf{0} & V_{2}   \\
  \end{array}
\right]
\cdot
\left[
  \begin{array}{cc}
    W'_{1} & \mathbf{0}   \\
    \mathbf{0} & W'_{2}   \\
  \end{array}
\right].
\end{eqnarray}

From equation \eqref{eq30}, $V_1$ and $V_2$ in equation \eqref{eq31} can be expressed as
\begin{eqnarray}\label{eq32}
\begin{split}
V_k=\prod_{i=1}^{\lfloor\frac{n_{k}}{2}\rfloor}
\textrm{exp}(-\textrm{i}\sigma_{x_{n_{k}}}^{i,\lfloor\frac{n_{k}}{2}\rfloor+i} \otimes D_{m}^{(k,i)}),\;\;
k\in\{1,2\},
\end{split}
\end{eqnarray}
where $D_m^{(k,i)}$ is a real diagonal matrix.
One can simply verify that
\begin{eqnarray}\label{eq33}
\begin{split}
&\left[
  \begin{array}{cc}
    V_{1} & \mathbf{0}   \\
    \mathbf{0} & V_{2}   \\
  \end{array}
\right]=
\left[
  \begin{array}{cc}
    V_{1} & \mathbf{0}   \\
    \mathbf{0} & I_{n_2}   \\
  \end{array}
\right]
\cdot
\left[
  \begin{array}{cc}
    I_{n_1} & \mathbf{0}   \\
    \mathbf{0} & V_{2}   \\
  \end{array}
\right],\\[2mm]
&\left[
  \begin{array}{cc}
    V_{1} & \mathbf{0}   \\
    \mathbf{0} & I_{n_2}   \\
  \end{array}
\right]=
\prod_{i=1}^{\lfloor\frac{n_{1}}{2}\rfloor}
\textrm{exp}(-\textrm{i}\sigma_{x_{n}}^{i,\lfloor\frac{n_{1}}{2}\rfloor+i} \otimes D_m^{(1,i)}),\\
&\left[
  \begin{array}{cc}
    I_{n_1} & \mathbf{0}   \\
    \mathbf{0} & V_{2}   \\
  \end{array}
\right]=
\prod_{i=1}^{\lfloor\frac{n_{2}}{2}\rfloor}
\textrm{exp}(-\textrm{i}\sigma_{x_{n}}^{n_1+i,n_1+\lfloor\frac{n_{2}}{2}\rfloor+i} \otimes D_m^{(2,i)}).
\end{split}
\end{eqnarray}
From equation \eqref{eq33}, the non-local operation $V_1\oplus V_2$ in equation \eqref{eq31} is equivalent to $\lfloor\frac{n_1}{2}\rfloor+\lfloor\frac{n_2}{2}\rfloor$ uniformly controlled $R_{x_{n}}$ gates with the control system $\mathcal{H}_{m}^{2}$. Hence, it can be implemented by local operations and $4(\lfloor\frac{n_1}{2}\rfloor+\lfloor\frac{n_2}{2}\rfloor)$ controlled-$X_{m}$ gates from figure \ref{Fig.2}(b) and figure \ref{Fig.4}.

As for $W_{1} \oplus W_{2}$ (and similarly for $W'_{1} \oplus W'_{2}$) in equation \eqref{eq31}, from equations \eqref{eq26} and \eqref{eq28}, we have
\begin{eqnarray}\label{eq34}
\begin{split}
\left[
  \begin{array}{cc}
    W_{1} & \mathbf{0}           \\
    \mathbf{0} & W_{2}           \\
  \end{array}
\right]
=
\left[
  \begin{array}{cccc}
    W_{11} & \mathbf{0} & \mathbf{0} & \mathbf{0}  \\
    \mathbf{0} & W_{12} & \mathbf{0} & \mathbf{0}  \\
    \mathbf{0} & \mathbf{0} & W_{21} & \mathbf{0}  \\
    \mathbf{0} & \mathbf{0} & \mathbf{0} & W_{22}  \\
  \end{array}
\right].
\end{split}
\end{eqnarray}
where $W_{11}\in U(\lfloor\frac{n_{1}}{2}\rfloor m)$, $W_{12}\in U((n_{1}-\lfloor\frac{n_{1}}{2}\rfloor)m)$,
$W_{21}\in U(\lfloor\frac{n_{2}}{2}\rfloor m)$, and $W_{22}\in U((n_{2}-\lfloor\frac{n_{2}}{2}\rfloor)m)$.

If $n_1=1$ and $n_2>1$ (i.e., $n_2=2$), it is only necessary to decompose $U_2$, in which $W_1=U_1$ and $W'_1=V_1=I_m$ in equation \eqref{eq31}.
For this case, since $U_1,W_{21},W_{22}\in U(m)$,
\begin{eqnarray}\label{eq35}
\begin{split}
\left[
  \begin{array}{cc}
    W_{1} & \mathbf{0}           \\
    \mathbf{0} & W_{2}           \\
  \end{array}
\right]
=
\left[
  \begin{array}{ccc}
    U_{1} & \mathbf{0} & \mathbf{0}   \\
    \mathbf{0} & W_{21} & \mathbf{0}   \\
    \mathbf{0} & \mathbf{0} & W_{22}   \\
  \end{array}
\right]
\end{split}
\end{eqnarray}
is a uniformly controlled unitary gate and its implementation is shown in figure \ref{Fig.3} and figure \ref{Fig.2}.
Thus, in this case, the synthesis of $X$ is completed.

\emph{Recursive decomposition of} $U(nm)$.--- 
If equation \eqref{eq34} is not a uniformly controlled unitary gate, the matrix in the right-hand side of equation \eqref{eq34} can be further decomposed following a similar argument as described above.
The decomposition terminates when all block matrices $W_{11}$, $W_{12}$, $W_{21}$, and $W_{22}$ reduce to a $m \times m$ matrix (i.e., equation \eqref{eq34} is a uniformly controlled unitary gate with control system $\mathcal{H}_{n}^{1}$).
Generally, the complete process requires $d=\lceil \log_{2}n \rceil$ steps of decomposition:
\begin{eqnarray}\label{eq36}
\begin{split}
U(nm) \xrightarrow{1} &\; U(\textstyle\lfloor\frac{n}{2}\rfloor m) \oplus U((n-\lfloor\frac{n}{2}\rfloor) m)\\
\xrightarrow{2} &\; 
\big[U(\textstyle\lfloor\frac{n_1}{2}\rfloor m) \oplus U((n_1-\lfloor\frac{n_1}{2}\rfloor) m)\big]
\oplus  
\big[U(\textstyle\lfloor\frac{n_2}{2}\rfloor m) \oplus U((n_2-\lfloor\frac{n_2}{2}\rfloor) m)\big]\\
&\cdots\\
\xrightarrow{d} &\; \textstyle\bigoplus_{m}U(n),
\end{split}
\end{eqnarray}
where $n_1=\lfloor\frac{n}{2}\rfloor$, $n_2=n-\lfloor\frac{n}{2}\rfloor$, and $\bigoplus_{m}U(n)$ denotes the direct sum of $m$ copies of $U(n)$.

\emph{Example}.---
When $n=5$, for any $X \in U(5m)$, its quantum circuit obtained after 3 steps of decomposition is shown in figure \ref{Fig.6}.
This process is given in Appendix \ref{Appendix.B}.

As shown in figure \ref{Fig.6}, we write $V^{(k)}$ to denote the non-local operation in the form of the product of uniformly controlled $R_{x_{n}}$ gates with the control system $\mathcal{H}_{m}^{2}$ obtained by the $k$-th decomposition.
In general, the final quantum circuit of $X\in U(nm)$ consists of $2^d$ uniformly controlled unitary gates with the control system $\mathcal{H}_{n}^{1}$ and $2^{d}-1$ non-local operations  $\{V^{(1)},\ldots,V^{(d)}\}$.
Finally, together with figures \ref{Fig.2}-\ref{Fig.4} and figure \ref{Fig.1}(c), $X$ can be synthesized by controlled-$X_{m}$ gates $\textrm{C}_{n}(X_{m})$ and local operations.

\begin{figure*}[htbp]
  \centering
  \includegraphics[width=15cm]{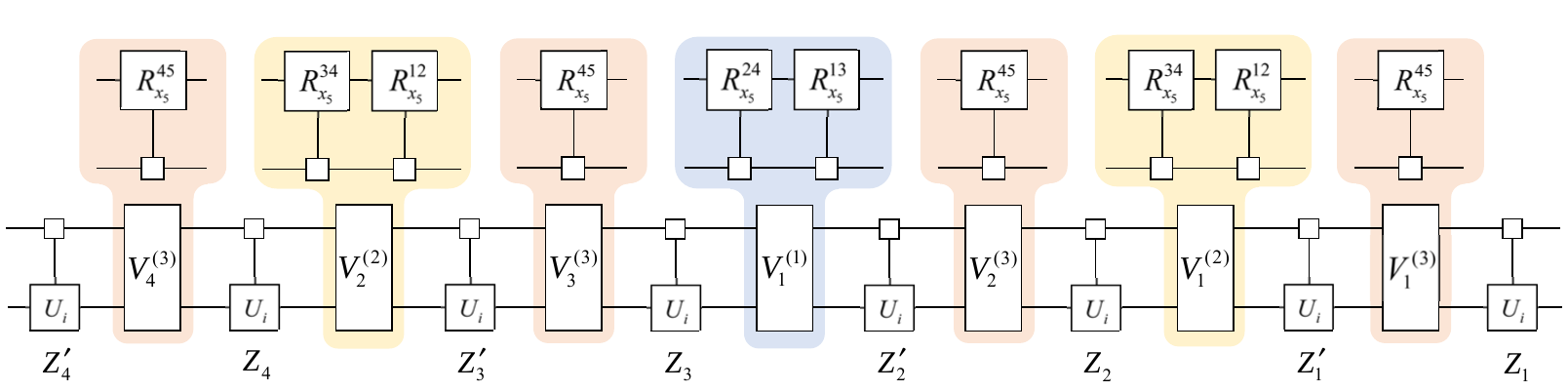}
  \caption{Quantum circuit for general unitary operations on $\mathcal{H}_{5}^{1}\otimes\mathcal{H}_{m}^{2}$. Appendix \ref{Appendix.A} provides a detailed explanation.}\label{Fig.6}
\end{figure*}

\subsection{Further simplification of quantum circuits for general quNit-quMit gates}\label{sec4.2}
Note that the constructed quantum circuit of $X\in U(nm)$ is implemented by an arrangement of $ZVZV\ldots ZVZ$,
where $Z$ denotes a uniformly controlled unitary gate and $V$ denotes the product of uniformly controlled $R_{x_{n}}$ gates in the circuit.
An example is shown in figure \ref{Fig.6}. 
From figure \ref{Fig.3}, we know that $Z$ consists of $n-1$ controlled unitary gates.
Applying the commutativity of matrices, we can further reduce the number of controlled unitary gates generated by the implementation of $Z$.

In particular, when $n$ is odd, it follows from equations \eqref{eq27} and \eqref{eq30} that
\begin{eqnarray}\label{eq37}
\begin{split}
V
=\left[
  \begin{array}{cc}
    \widetilde{V} & \mathbf{0}  \\
    \mathbf{0} & I_{m}       \\
  \end{array}
\right]
=\prod_{i=1}^{\frac{n-1}{2}}
\textrm{exp}(-\textrm{i}\sigma_{x_n}^{i,\frac{n-1}{2}+i} \otimes D_m^{(i)}),
\end{split}
\end{eqnarray}
where $\widetilde{V}$ is a certain $(n-1)m \times (n-1)m$ unitary matrix. 
Indeed, it can be calculated that
\begin{eqnarray}\label{eq38}
\begin{split}
\widetilde{V}=\left[
  \begin{array}{cc}
    C  & -\textrm{i}S \\
    -\textrm{i}S & C       \\
  \end{array}
\right]
=\prod_{i=1}^{\frac{n-1}{2}}
\textrm{exp}(-\textrm{i}\sigma_{x_{n-1}}^{i,\frac{n-1}{2}+i} \otimes D_m^{(i)}).
\end{split}
\end{eqnarray}
Hence, from equation \eqref{eq6}, it holds that 
\begin{eqnarray}\label{eq39}
\begin{split}
\textrm{C}_{n}(U) \cdot V=V \cdot \textrm{C}_{n}(U),
\end{split}
\end{eqnarray}
where $U$ is a unitary operator on $\mathcal{H}_{m}^{2}$.
From equation \eqref{eq39}, $\textrm{C}_{n}(U)$ generated by $Z$ can be transferred to the right side of $V$, and then combined with $Z'$ in the next block to form a new uniformly controlled unitary gate.
Thus, we can eliminate one controlled unitary gate, which reduces two controlled-$X_{m}$ gates in the circuit.

Here we still assume $n=5$ to explain how to eliminate some controlled unitary gates in figure \ref{Fig.6}.
From equations \eqref{eqA2}, \eqref{eqA5}, and \eqref{eqA10}, it follows that
\begin{eqnarray}\label{eq40}
V^{(i)}=\left[
  \begin{array}{cc}
    \widetilde{V}^{(i)} & \mathbf{0}           \\
    \mathbf{0} & I_{m}       \\
  \end{array}
\right],\;
V^{(3)}=\left[
  \begin{array}{cc}
     I_{3m} &\mathbf{0}      \\
    \mathbf{0} & \widetilde{V}^{(3)}           \\
  \end{array}
\right],
\end{eqnarray}
where $i\in\{1,2\}$ and
\begin{eqnarray}\label{eq41}
\begin{split}
\widetilde{V}^{(1)}&=\textrm{exp}(-\textrm{i}\sigma_{x_4}^{13} \otimes D_m^{(1)})\cdot\textrm{exp}(-\textrm{i}\sigma_{x_4}^{24} \otimes D_m^{(2)}),\\[1mm]
\widetilde{V}^{(2)}&=\textrm{exp}(-\textrm{i}\sigma_{x_4}^{12} \otimes D_m^{(3)})\cdot\textrm{exp}(-\textrm{i}\sigma_{x_4}^{34} \otimes D_m^{(4)}),\\[1mm]
\widetilde{V}^{(3)}&=\textrm{exp}(-\textrm{i}\sigma_{x_2}^{12} \otimes D_m^{(5)}).
\end{split}
\end{eqnarray}
Thus, it holds that
\begin{eqnarray}\label{eq42}
\begin{split}
\textrm{C}_{5}(U) \cdot V^{(i)}&=V^{(i)} \cdot \textrm{C}_{5}(U),\;\;i\in\{1,2\},\\
\textrm{C}_{j}(U) \cdot V^{(3)}&=V^{(3)} \cdot \textrm{C}_{j}(U),\;\;j\in\{1,2,3\}.
\end{split}
\end{eqnarray}
In figure \ref{Fig.6}, equation \eqref{eq42} implies that $\textrm{C}_{5}(U)$ (resp. $\textrm{C}_{j}(U)$), which arises from the synthesis of $Z_{k}$ $(k\in\{2,3,4\})$ (resp. $Z'_{l}$ $(l\in\{1,2,3,4\})$) on the left of $V^{(i)}$ (resp. $V^{(3)}$), can be absorbed by $Z'_{k-1}$ (resp. $Z_{l}$) on the right of $V^{(i)}$ (resp. $V^{(3)}$).
Thus, for the implementation of $Z_{k}$ $(k\in\{2,3,4\})$, we can omit one controlled unitary gate; $Z'_{l}$ $(l\in\{1,2,3,4\})$ can omit three controlled unitary gates.
Therefore, 15 controlled unitary gates in figure \ref{Fig.6} may be omitted.

\section{Quantum gate count}\label{sec5}
Below, we calculate the number of controlled-$X_{m}$ gates required to synthesize a general unitary operation $X\in U(nm)$ on $\mathcal{H}_{n}^{1}\otimes\mathcal{H}_{m}^{2}$ by this scheme.
For this purpose, it is only sufficient to identify the number of uniformly controlled $R_{x_{n}}$ gates with the control system $\mathcal{H}_{m}^{2}$ required for the synthesis of $V^{(k)}$, where $k\in\{1,\ldots,d\}$, and the number of controlled unitary gates that can be eliminated.

Given a positive integer $n$, let $d=\lceil \log_{2}n \rceil$ and $n(1,1)=n$.
Fix an integer $k\in\{2,\ldots,d\}$.
For $i\in\{1,\ldots,2^{k-1}\}$, we define 
\begin{eqnarray}\label{eq43}
\begin{split}
n(k,i)=\left\{
\begin{array}{cc}
\lfloor n(k-1,\frac{i+1}{2}) \rfloor,       & \textrm{if}\;i\;\textrm{is}\;\textrm{odd},   \\[2mm]
 n(k-1,\frac{i}{2})-\lfloor n(k-1,\frac{i}{2}) \rfloor,  & \textrm{if}\;i\;\textrm{is}\;\textrm{even}.     \\
\end{array}
\right.
\end{split}
\end{eqnarray}
Then we get the decomposition $n=\sum^{2^{k-1}}_{i=1} n(k,i)$ for each $k\in\{1,2,\ldots,d\}$.
For example, 
\begin{eqnarray}\label{eq44}
\begin{split}
5&=2+3\\
&=(1+1)+(1+2).
\end{split}
\end{eqnarray}
where one has $5(1,1)=5$, $5(2,1)=2$, $5(2,2)=3$, $5(3,1)=5(3,2)=5(3,3)=1$, and $5(3,4)=2$.

For $k\in\{1,\ldots,d\}$, let $N^{(k)}_n$ denote the number of odd integers in the set $\{n(k,1),\ldots,n(k,2^{k-1})\}$, i.e.,
\begin{eqnarray}\label{eq45}
\begin{split}
N^{(k)}_n=\sum_{i=1}^{2^{k-1}}\big[n(k,i)\;\textrm{mod}\;2\big].
\end{split}
\end{eqnarray}
From the decomposition given in equation \eqref{eq26}, $N^{(k)}_n$ means the number of identity matrices $I_m$ in the diagonal blocks of $V^{(k)}$.
For the case $n=5$, we have $N^{(1)}_5=N^{(2)}_5=1$ and $N^{(3)}_5=3$.
Comparing with equation \eqref{eq40}, it can be observed that $N^{(1)}_5$, $N^{(2)}_5$, and $N^{(3)}_5$ indeed correspond to the number of the identity matrices $I_m$ in the diagonal blocks of $V^{(1)}$, $V^{(2)}$, and $V^{(3)}$, respectively.

Based on the method described in section \ref{sec4.2}, the number of controlled unitary gates that can be eliminated is equal to the number of identity matrices $I_m$ in the diagonal blocks of $V^{(k)}$.
Therefore, the number of controlled unitary gates that can be eliminated is $\sum_{k=1}^{d}2^{k-1}N^{(k)}_n$ by the fact that there are $2^{k-1}$ non-local operations $V^{(k)}$ in the quantum circuit of $X$.
Moreover, the quantum circuit of $V^{(k)}$ consists of $\frac{n-N^{(k)}_{n}}{2}$ uniformly controlled $R_{x_{n}}$ gates with the control system $\mathcal{H}_{m}^{2}$.

There are $2^{d}$ uniformly controlled unitary gates with the control system $\mathcal{H}_{n}^{1}$ in the quantum circuit of $X$.
Thus, the number of controlled-$X_{m}$ gates required to synthesize a general quNit-quMit gate $X$ is at most
\begin{eqnarray}\label{eq46}
\begin{split}
(2n-1)2^{\lceil \log_{2}n \rceil+1}-2n-\sum_{k=1}^{\lceil \log_{2}n \rceil}2^{k+1}N^{(k)}_n.
\end{split}
\end{eqnarray}
Since $2^{\lceil \log_{2}n \rceil+1}=O(n)$, the quantum circuit of $X\in U(nm)$ is implemented by $O(n^2)$ controlled-$X_{m}$ gates.
Appendix \ref{Appendix.C} gives a code to calculate the number of CINC gates.
Table \ref{Table1} shows a comparison of the number of quantum gates with the previous synthesis schemes.

\begin{table}[htbp]
\centering
\caption{Comparison of the number of imprimitive gates required to synthesize a general unitary operation on $\mathcal{H}_{n}^{1}\otimes\mathcal{H}_{n}^{2}$ by several schemes.}\label{Table1}
\begin{tabular}{cccccccc}
\hline
Synthesis  & imprimitive &\multicolumn{6}{c}{Gate count for $3\leq n \leq 8$} \\ \cline{3-8}
Algorithm  &    gates    & 3  & 4   &5    & 6   & 7    & 8 \\ \hline
QSD \cite{Di2015}       & GCX     & 26 & 90  & 176 & 355 & 618  & 980 \\
QSD \cite{Li-Wen-Dong}  & CDNOT   & \textendash & 60 & \textendash & \textendash & \textendash & \textendash \\
QR \cite{Brennen2006}   & CINC \& CINC$^{-1}$ & 78 & 220 & 495 & 996 & 1708 & 2808 \\
CSD \cite{Nakajima2009} & CINC \& CINC$^{-1}$ & 36 & 72  & 280 & 420 & 588  & 784  \\
Our                     & CINC  & \textbf{19} & \textbf{48} & \textbf{74}  & \textbf{116} & \textbf{166} & \textbf{224}\\
\hline
\end{tabular}
\end{table}

Moreover, for $\mathcal{H}_{n}^{1}\otimes\mathcal{H}_{n}^{2}$, Ref. \cite{Brennen2005} shows that $O(n^4)$ controlled-phase gates $e^{\textrm{i}\pi|n\rangle\langle n| \otimes |n\rangle\langle n|}$ are required to implement a general unitary gate on $\mathcal{H}_{n}^{1}\otimes\mathcal{H}_{n}^{2}$.
From equation \eqref{eq16} in Ref. \cite{Brennen2005}, a controlled-$X_{n}$ gate can be implemented by $n-1$ controlled-phase gates.
Thus, our protocol requires at most $O(n^3)$ controlled-$X_{n}$ gates to implement a two-quNit gate,
which establishes a better upper bound than the $O(n^4)$ complexity in Ref. \cite{Brennen2005}.

\section{Conclusion}\label{sec6}
We have presented a recursive algorithm to exactly synthesize an arbitrary unitary operation on $\mathcal{H}_{n}^{1}\otimes\mathcal{H}_{m}^{2}$.
The constructed quantum circuit for a general unitary operation consists of the controlled-$X_{m}$ gate $\textrm{C}_{n}(X_{m})$ and local gates without ancillary quantum systems.
The control states of all controlled-$X_{m}$ gates in the quantum circuit are located on the system $\mathcal{H}_{n}^{1}$.
Since an arbitrary multi-quNit gate can be exactly simulated by two-quNit gates Ref. \cite{Bullock2005}, our scheme shows that $\textrm{C}_{n}(X_{m})$ and local gates are universal for HD quantum computation.

The number of $\textrm{C}_{n}(X_{m})$ in the circuit is given by equation \eqref{eq46}, which is determined by the dimension of $\mathcal{H}_{n}^{1}$ and independent of $\mathcal{H}_{m}^{2}$.
Combined with the trick of ignoring controlled unitary gates, the number of $\textrm{C}_{n}(X_{m})$ in the circuit is greatly reduced compared with previous works.
As shown in table \ref{Table1}, the result shows that the complexity of our circuit is the lowest currently known.
Our scheme is not limited to a particular physical system, which has potential for HD quantum computation.
Future research directions include extending the quantum circuit to multi-component systems (involving more than two systems) and designing experimental schemes for this quantum circuit.


\section*{Funding} \par

This work is supported by the National Natural Science Foundation of China under Grant No. 62371038.

\section*{Author contributions} \par

Both authors had equal contributions to the paper.

\section*{Data availability} \par

All data that support the findings of this study are included with in the article (and any supplementary files)


\section*{Appendix}

\appendix

\section{Experimental implementation of controlled increment gates}\label{Appendix.A}

We now present an experimental scheme to implement the controlled-$X_{m}$ gate $\textrm{C}_{n}(X_{m})$ following the method in Ref. \cite{Gao2020}.
Here we take $\textrm{C}_{3}(X_{3})$ on $\mathcal{H}_{3}^{1}\otimes\mathcal{H}_{3}^{2}$ as an example, while the generalization to arbitrary dimensions follows similarly (see also Supplementary Material of Ref. \cite{Gao2020}).
We consider the orbital angular momentum (OAM) degree of freedom of photons as the quantum states, which means that $|l\rangle$ denotes a single photon with OAM value $l$.

\begin{figure}[htbp]
  \centering
  \includegraphics[width=8.5cm]{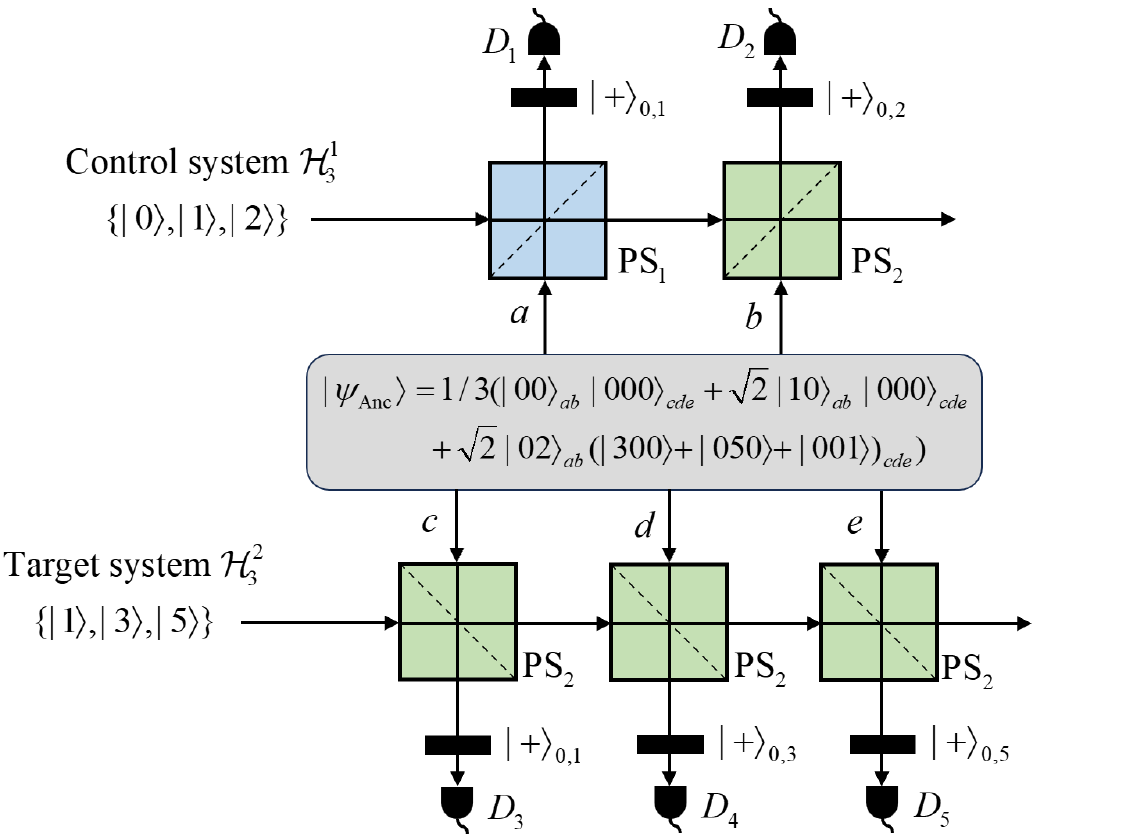}
  \caption{Experimental setup of controlled increment gates for $\mathcal{H}_{3}^{1}\otimes\mathcal{H}_{3}^{2}$. Here $\textrm{PS}_1$ denotes a parity sorter, which transmits even modes $\{|0\rangle,|2\rangle,\ldots\}$ and reflects odd modes $\{|1\rangle,|3\rangle,\ldots\}$.
  $\textrm{PS}_2$ denotes a second-order parity sorter, which transmits modes $|2k\times2\rangle$ ($k=0,1,2,\ldots$) and reflects modes $|(2k-1)\times2\rangle$.
  If the input is an odd mode, then $\textrm{PS}_2$ transmits the photon with a probability of $\frac{1}{2}$ and reflects the photon with equal probability.
  $|+\rangle_{0,k}$ denotes a mode filter, which projects the photons into the subspace $\{|0\rangle,|k\rangle\}$.
  $D_i$ denotes a detector.}\label{Fig7}
\end{figure}

The experimental setup for implementing $\textrm{C}_{3}(X_{3})$ is shown in figure \ref{Fig7}.
The implementation requires an ancillary quantum state 
\begin{eqnarray}\label{eq47}
\begin{split}
|\psi_{\textrm{Anc}}\rangle=&\frac{1}{3}|00\rangle_{ab}|000\rangle_{cde}
+\frac{\sqrt{2}}{3}|10\rangle_{ab}|000\rangle_{cde}+\frac{\sqrt{2}}{3}|02\rangle_{ab}(|300\rangle+|050\rangle+|001\rangle)_{cde}.
\end{split}
\end{eqnarray}
The subscripts $abcde$ here mean the photons are in the corresponding paths as shown in figure \ref{Fig7}.
The desired quantum operation is heralded by the simultaneous clicks of all detectors and relies on post-selection.
The only difference between this setup and figure 2 in Ref. \cite{Gao2020} is the ancillary quantum state.

Specifically, if the state of the control system $\mathcal{H}_{3}^{1}$ is $|2\rangle$, the two detectors $D_1$ and $D_2$ click and the output state is still $|2\rangle$, only when the photons state in the paths $ab$ is $|02\rangle_{ab}$.
In other cases, either the detectors $D_1$ and $D_2$ do not all fire, or there is no photon in the output path (see also figure 2 of Ref. \cite{Gao2020}).
Thus, if the input state is $|2\rangle$ and the detectors $D_1$ and $D_2$ click, then $|\psi_{\textrm{Anc}}\rangle$ will collapse into $\frac{1}{\sqrt{3}}(|300\rangle+|050\rangle+|001\rangle)_{cde}$ with a probability of $2/3$.
In this case, if detectors $D_3$-$D_5$ all click, the state of the target system $\mathcal{H}_{3}^{2}$ follows the transformations: $|1\rangle\rightarrow|3\rangle,|3\rangle\rightarrow|5\rangle,|5\rangle\rightarrow|1\rangle$ with a probability of $1/(2^4\times3)$ (see figure 1 of Ref. \cite{Gao2020}).

If the state of $\mathcal{H}_{3}^{1}$ is $|0\rangle$ (resp. $|1\rangle$), only $|00\rangle_{ab}$ (resp. $|10\rangle_{ab}$) can cause both detectors $D_1$ and $D_2$ to click and the output state to be $|0\rangle$ (resp. $|1\rangle$).
Thus, when the input state is $|0\rangle$ or $|1\rangle$ and detectors $D_1$ and $D_2$ click, $|\psi_{\textrm{Anc}}\rangle$ collapses into $|000\rangle_{cde}$ with a probability of $1/9$.
The state $|000\rangle_{cde}$ activates detectors $D_3$-$D_5$ while leaving the output state of the target system unchanged with a probability of $1/2^3$.

Therefore, when detectors $D_1$-$D_5$ click, the quantum gate $\textrm{C}_{3}(X_{3})$ is successfully implemented.
The success probability of the implementation for $\textrm{C}_{3}(X_{3})$ is $1/(2^3\times9)=1/72$, regardless of the input mode of the control system.

\section{Example: Decomposition of elements in \emph{U}(5\emph{m})}\label{Appendix.B}
Assuming that $X \in U(5m)$, we will give the decomposition of $X$ through the method described in section \ref{sec4.1}.

Step 1: We use equation \eqref{eq26} to decompose $X$ into
\begin{eqnarray}\label{eqA1}
\begin{split}
X=U \cdot V^{(1)} \cdot U'.
\end{split}
\end{eqnarray}
From equation \eqref{eq30}, we have 
\begin{eqnarray}\label{eqA2}
\begin{split}
V^{(1)}=
\textrm{exp}(-\textrm{i}\sigma_{x_5}^{13} \otimes D_m^{(1)})\cdot\textrm{exp}(-\textrm{i}\sigma_{x_5}^{24} \otimes D_m^{(2)}),
\end{split}
\end{eqnarray}
where $D_m^{(1)}$ and $D_m^{(2)}$ are real diagonal matrices. 
For $U$, it follows from equation \eqref{eq25} that
\begin{eqnarray}\label{eqA3}
\begin{split}
U=\left[
  \begin{array}{cc}
    U_{1} & \mathbf{0}           \\
    \mathbf{0} & U_{2}           \\
  \end{array}
\right],
\end{split}
\end{eqnarray}
where $U_1 \in U(2m)$ and $U_2 \in U(3m)$.
The treatment of $U'$ parallels that of $U$.

Step 2: We then use equation \eqref{eq26} to decompose $U_1$ and $U_2$.
From equation \eqref{eq31}, we get
\begin{eqnarray}\label{eqA4}
\begin{split}
U=W \cdot V^{(2)} \cdot W'.
\end{split}
\end{eqnarray}
By equation \eqref{eq33}, $V^{(2)}$ can be expressed as
\begin{eqnarray}\label{eqA5}
\begin{split}
V^{(2)}=
\textrm{exp}(-\textrm{i}\sigma_{x_5}^{12} \otimes D_m^{(3)})\cdot\textrm{exp}(-\textrm{i}\sigma_{x_5}^{34} \otimes D_m^{(4)}),
\end{split}
\end{eqnarray}
where $D_m^{(3)}$ and $D_m^{(4)}$ are real diagonal matrices.
Moreover, from equation \eqref{eq34}, one has
\begin{eqnarray}\label{eqA6}
\begin{split}
W=
\left[
  \begin{array}{cccc}
    W_{11} & \mathbf{0} & \mathbf{0} & \mathbf{0}  \\
    \mathbf{0} & W_{12} & \mathbf{0} & \mathbf{0}  \\
    \mathbf{0} & \mathbf{0} & W_{21} & \mathbf{0}  \\
    \mathbf{0} & \mathbf{0} & \mathbf{0} & W_{22}  \\
  \end{array}
\right],
\end{split}
\end{eqnarray}
where $W_{11},W_{12},W_{21} \in U(m)$ and $W_{22} \in U(2m)$.
An analogous discussion holds for $W'$.

Step 3: Finally, we just need to decompose $W_{22}$.
Using equation \eqref{eq26} for $W_{22}$, we get 
\begin{eqnarray}\label{eqA7}
\begin{split}
W_{22}=
\left[
  \begin{array}{cc}
    \widetilde{Z}_{1} & \mathbf{0}           \\
    \mathbf{0} & \widetilde{Z}_{2}           \\
  \end{array}
\right]
\cdot
\textrm{exp}(-\textrm{i}\sigma_{x_2}^{12} \otimes D_m^{(5)})
\cdot
\left[
  \begin{array}{cc}
    \widetilde{Z}'_{1} & \mathbf{0}           \\
    \mathbf{0} & \widetilde{Z}'_{2}           \\
  \end{array}
\right],
\end{split}
\end{eqnarray}
where $\{\widetilde{Z}_{1},\widetilde{Z}'_{1},\widetilde{Z}_{2},\widetilde{Z}'_{2}\} \subset U(m)$.
Substituting equation \eqref{eqA7} into equation \eqref{eqA6}, it immediately follows that
\begin{eqnarray}\label{eqA8}
\begin{split}
W=Z \cdot V^{(3)} \cdot Z',
\end{split}
\end{eqnarray}
where 
\begin{eqnarray}\label{eqA9}
\begin{split}
Z=\left[
  \begin{array}{ccccc}
    W_{11} & \mathbf{0} & \mathbf{0} & \mathbf{0} & \mathbf{0} \\
    \mathbf{0} & W_{12} & \mathbf{0} & \mathbf{0} & \mathbf{0} \\
    \mathbf{0} & \mathbf{0} & W_{21} & \mathbf{0} & \mathbf{0} \\
    \mathbf{0} & \mathbf{0} & \mathbf{0} & \widetilde{Z}_{1} & \mathbf{0} \\
    \mathbf{0} & \mathbf{0} & \mathbf{0} & \mathbf{0} & \widetilde{Z}_{2} \\
  \end{array}
\right],\;\;
Z'=\left[
  \begin{array}{ccc}
     I_{3m} & \mathbf{0} & \mathbf{0} \\
     \mathbf{0} & \widetilde{Z}'_{1} & \mathbf{0} \\
     \mathbf{0} & \mathbf{0} & \widetilde{Z}'_{2} \\
  \end{array}
\right],
\end{split}
\end{eqnarray}
and 
\begin{eqnarray}\label{eqA10}
\begin{split}
V^{(3)}&=
\left[
  \begin{array}{cc}
    I_{3m} &  \mathbf{0} \\
      \mathbf{0}  & \textrm{exp}(-\textrm{i}\sigma_{x_2}^{12} \otimes D_m^{(5)}) \\
  \end{array}
\right]\\[1mm]
&=\textrm{exp}(-\textrm{i}\sigma_{x_5}^{45} \otimes D_m^{(5)}).
\end{split}
\end{eqnarray}
Note that $Z$ and $Z'$ are uniformly controlled unitary gates with the control system $\mathcal{H}_{5}^{1}$.

Hence, from equations \eqref{eqA1}, \eqref{eqA4}, and \eqref{eqA8}, we can obtain
\begin{eqnarray}\label{eqA11}
\begin{split}
X=&\;U \cdot V^{(1)} \cdot U'\\
=&\;W_1 \cdot V^{(2)}_{1} \cdot W'_1 \cdot V^{(1)} \cdot W_2 \cdot V^{(2)}_{2} \cdot W'_2\\
=&\;(Z_1 \cdot V^{(3)}_{1} \cdot Z'_1) \cdot V^{(2)}_{1} \cdot (Z_2 \cdot V^{(3)}_{2} \cdot Z'_2) \cdot V^{(1)} 
\\&\cdot (Z_3 \cdot V^{(3)}_{3} \cdot Z'_3) \cdot V^{(2)}_{2} \cdot (Z_4 \cdot V^{(3)}_{4} \cdot Z'_4).
\end{split}
\end{eqnarray}
Here $Z_i$ and $Z'_i$ are uniformly controlled unitary gates with the control system $\mathcal{H}_{5}^{1}$.
Based on equation \eqref{eqA11}, we can get the quantum circuit of $X$, as shown in figure \ref{Fig.6}.

\section{Algorithm for the CINC gate counts}\label{Appendix.C}

\begin{figure}[htbp]
  \centering
  \includegraphics[width=10cm]{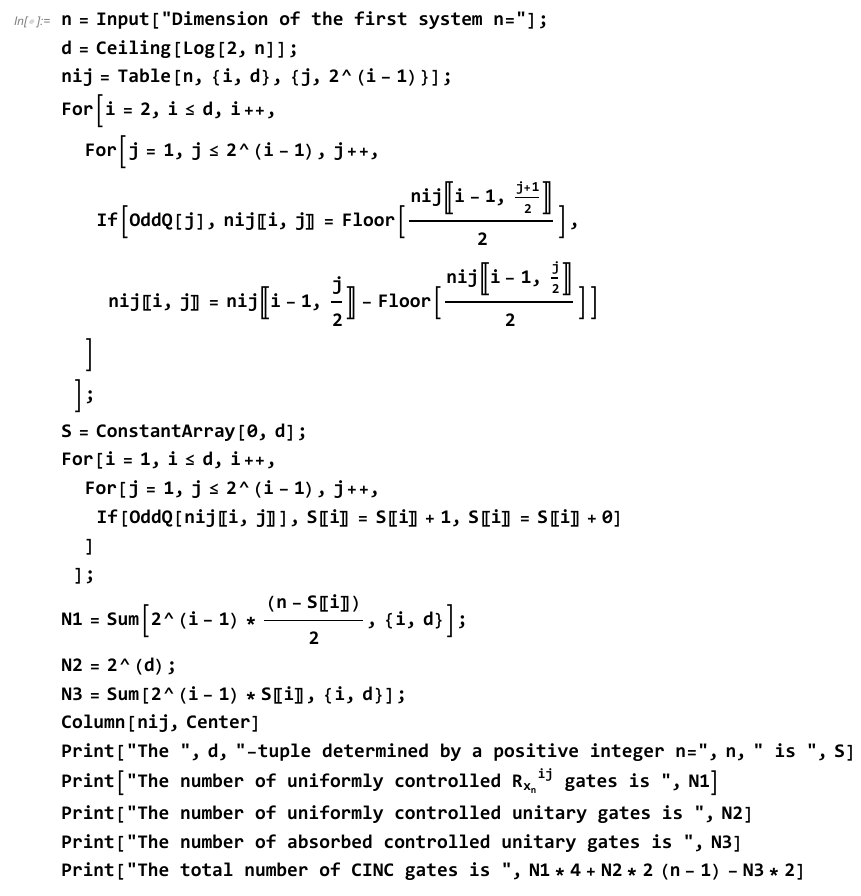}
  \caption{Code for calculating CINC gate counts in Wolfram Mathematica.}\label{Fig8}
\end{figure}

Given an arbitrary dimension $n$, the Wolfram code to calculate the number of CINC required to accurately implement a general quNit-quMit gate based on our scheme is shown in figure \ref{Fig8}.

\end{document}